\documentclass[aps,prb,twocolumn,showpacs,superscriptaddress, longbibliography]{revtex4-2}

\usepackage{xcolor, graphicx}
\usepackage{amssymb,amsmath, amsfonts, bm}
\usepackage{float}
\usepackage{dcolumn}
\usepackage[none]{hyphenat} 
\usepackage{tabularx}

\usepackage{newtxtext,newtxmath}

\usepackage{verbatim}  

\begin{document}

\title{High-temperature quantum spin Hall states in buckled III-V-monolayer/SiO$_{2}$}
\author{Yunyouyou Xia}
\email{These two authors contributed equally to this work.}
\affiliation{School of Physical Science and Technology, ShanghaiTech University, Shanghai 200031, China}
\affiliation{\mbox{ShanghaiTech Laboratory for Topological Physics, ShanghaiTech University, Shanghai 200031, China}}
\author{Suhua Jin}
\email{These two authors contributed equally to this work.}
\affiliation{School of Physical Science and Technology, ShanghaiTech University, Shanghai 200031, China}
\author{Werner Hanke}
\affiliation{Institut f\"{u}r Theoretische Physik und Astrophysik, Universit\"{a}t W\"{u}rzburg, D-97074 W\"{u}rzburg, Germany}
\affiliation{W\"{u}rzburg-Dresden Cluster of Excellence ct.qmat, Universit\"{a}t W\"{u}rzburg, D-97074 W\"{u}rzburg, Germany}
\author{Ralph Claessen}
\affiliation{Physikalisches Institut, Universit\"{a}t W\"{u}rzburg, D-97074 W\"{u}rzburg, Germany}
\affiliation{W\"{u}rzburg-Dresden Cluster of Excellence ct.qmat, Universit\"{a}t W\"{u}rzburg, D-97074 W\"{u}rzburg, Germany}
\author{Gang Li}
\email{ligang@shanghaitech.edu.cn}
\affiliation{School of Physical Science and Technology, ShanghaiTech University, Shanghai 201210, China}
\affiliation{\mbox{ShanghaiTech Laboratory for Topological Physics, ShanghaiTech University, Shanghai 201210, China}}

\begin{abstract}
After establishing the fundamental understanding and the high throughput topological characterization of nearly all inorganic three-dimensional materials,  the general interest and the demand of functional applications drive the research of topological insulators to the exploration of systems with a more robust topological nature and fewer fabrication challenges.  
The successful demonstration of the room-temperature quantum spin Hall (QSH) states in bismuthene/SiC(0001), thus, triggers the search of two-dimensional topological systems that are experimentally easy to access and of even larger topological gaps.
In this work, we propose a family of III-V honeycomb monolayers on SiO$_{2}$ to be the next generation of large gap QSH systems, based on which a spintronic device may potentially operate at room temperature due to its enlarged topological gap ($\sim$ 900 meV) as compared to bismuthene/SiC(0001). 
Fundamentally, this also realizes a band-inversion type QSH insulator that is distinct to the Kane-Mele type bismuthene/SiC(0001).
\end{abstract}

\maketitle

{\it{Introduction.}} -- 
Quantum spin Hall (QSH) insulators are considered as the key to the next generation of spintronics devices utilizing dissipationless edge spin currents. 
Yet, after initial suggestions in graphene~\cite{PhysRevLett.95.226801, PhysRevLett.95.146802} and experimental observation in HgTe quantum wells~\cite{hgte1, hgte2}, most proposed QSH systems are hindered from further device applications due to either minute topological gaps or poor chemical stability. This leads to vast attempts devoted to construct experimentally feasible high-temperature QSH systems. The topological bulk gaps are closely associated with the spin-orbit coupling (SOC) strength. Hence, in the frame of graphene-type systems, elemental honeycomb sheets containing heavy Xenes (X denotes element, e.g. silicene~\cite{PhysRevLett.107.076802}, germanene~\cite{PhysRevLett.107.076802}, stanene~\cite{PhysRevLett.111.136804} and bismuthene~\cite{PhysRevLett.107.136805, PhysRevB.83.121310, PhysRevLett.97.236805}) were predicted to possess sizable topological bulk gaps. Meanwhile, there is a chasm between free-standing models and realistic samples, the latter of which, have to be eventually placed or grown on substrates. Therefore, QSH systems used for spintronic devices must be designed in favor of experimental fabrications. In this context, substantial progress was achieved in bismuthene on SiC (0001)~\cite{Reis287, PhysRevB.98.165146}, possessing so-far the largest QSH gap ($\sim$ 800 meV) both experimentally probed and theoretically verified. In the present paradigm, the above challenges are coped by constructing a monolayer/substrate heterostructure, which offers a ``best of two worlds'' effect: the substrate not only stabilizes the QSH insulator, but also enlarges the topological gap via orbital filtering.

The successful fabrication of bismuthene/SiC(0001) significantly benefits from the progress made for preparing a smooth and defect-free SiC(0001) surface with hydrogen termination~\cite{doi:10.1021/acs.jpcc.6b01493}, which is a highly challenging task and is critical to the topological nature of the heterostructure. 
The surface of the substrate needs to be smooth enough in a large extent to host the flat bismuthene monolayer. 
Meanwhile, a second requirement further imposes that the bismuth honeycomb needs to selectively bond with the underlying silicons where the latter arranges in triangle. 
Consequently, not every silicon atom at the surface of SiC(0001) will be saturated and there is one silicon remaining in each hexagon to be hydrogenated. 
On the other hand, QSH Xenes combine the Kane-Mele mechanism~\cite{PhysRevLett.95.226801, PhysRevLett.95.146802} with a non-trivial property stemming from the $K$ point, at which momentum the substrate-induced Rashba band splitting is comparably sizable. 
Such a Rashba splitting does not cooperate with the topology, but reduces the band gap. 
Moreover, in bismuthene/SiC(0001), the conduction band minimum lies at $\Gamma$ rather than $K$. 
The indirect gap, together with the sizable Rashba splitting at $K$, result in a global gap smaller than what $p_{x/y}$ can intrinsically support. 
The success of the bismuthene/SiC work encourages the search for alternative QSH systems utilizing the same promotion mechanism which, however, are free of the two issues, i.e., indirect gap and large Rashba splitting.
Recently, the search extends to alloy-type inversion-asymmetric III-V binary monolayers~\cite{doi:10.1021/nl500206u, Li2015}. 
Including buckled III-Bi monolayer, the free-standing model of which was predicted to host QSH states with gaps up to around $560$ meV \cite{doi:10.1021/nl500206u}. To simulate bonding effects of substrates, hydrogenated~\cite{doi:10.1021/acs.nanolett.5b02293, PhysRevB.91.235306, C8RA00369F, Kim2016, doi:10.1021/acs.jpcc.5b07961}, halogenated~\cite{PhysRevB.91.235306, BARHOUMI2018171, doi:10.1021/nl504493d, Freitas_2016, Kim2016} and molecule-functionalized~\cite{Li2016, PhysRevMaterials.2.014005, doi:10.1063/1.5033999, PhysRevB.94.195424} III-V films were studied, and part thereof, are proposed to preserve both stabilities and non-trivial edge states. Moreover, the possible synthesis of III-V monolayers on Si (111)~\cite{doi:10.1021/acs.nanolett.5b02293, Yao2015, DENISOV2016105, PhysRevB.98.121404}, $h$-BN~\cite{Li2016, PhysRevMaterials.2.014005, doi:10.1063/1.5033999,BARHOUMI2018171} and SrTe (111)~\cite{PhysRevB.94.195424} substrates have been predicted.

\begin{figure*}[t]
\centering
\includegraphics[width=\linewidth]{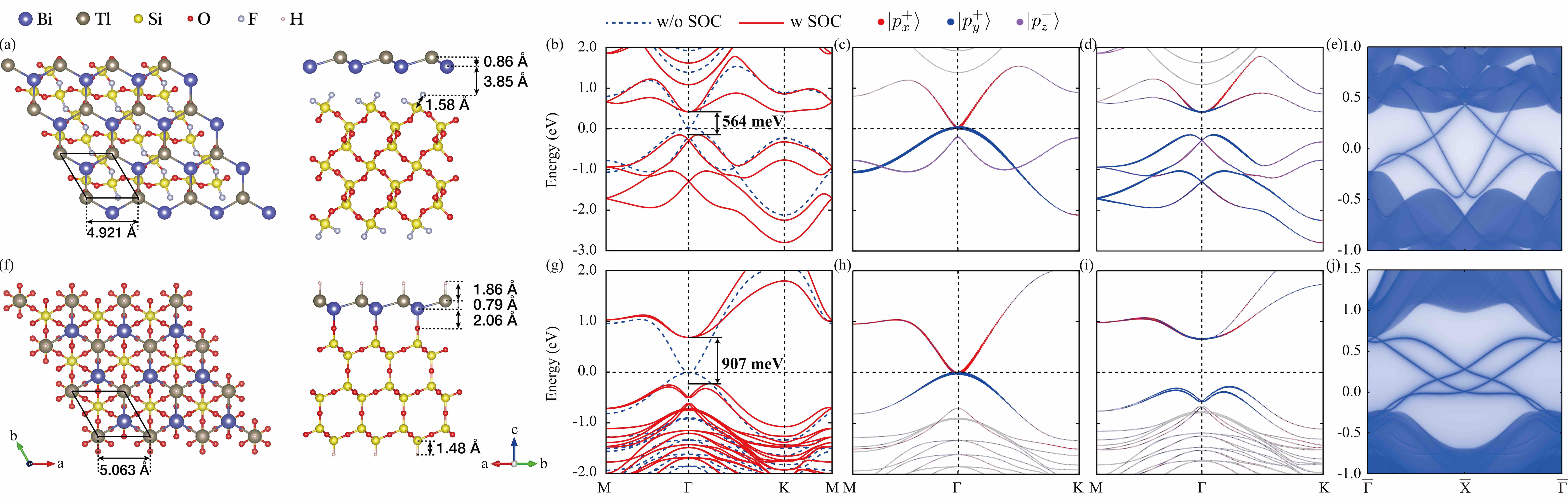}
\caption{(Color online) \textbf{III-V-monolayer/SiO$_2$.} (a) Crystalline structures of TlBi/$\alpha$-SiO$_2$ in top and side view. (b) Electronic structures of TlBi/$\alpha$-SiO$_2$. (c, d) Projection of Bloch bands on bonding/anti-bonding states without (c) and with (d) the SOC of TlBi/$\alpha$-SiO$_2$. (e) The zigzag edge states of TlBi/$\alpha$-SiO$_2$. (f-j) The same plot for TlBi/$\beta$-SiO$_2$.}
\label{Fig1}
\end{figure*}

Unfortunately, none of the above-mentioned attempts achieves the demand for a balance between a large topological gap and a high experimental feasibility, both of which are crucial for device application. 
In this article, we propose III-V honeycomb monolayers on SiO$_{2}$ substrate to be a new platform for large-gap QSH states, which likely renders less fabrication challenge than bismuthene/SiC(0001).

The two silica polymorphs support III-V monolayers in different ways, to some extent, releasing the critical bonding requirement in bismuthene/SiC(0001).
Furthermore, they realize a band-inversion type QSH phase in contrast to the Kane-Mele bismuthene/SiC(0001). Thus, together with bismuthene/SiC, our work completes the implementation of the fascinating ``best of two worlds" for both topological mechanisms in honeycomb heterostructures.

{\it{III-V-monolayer/SiO$_{2}$.}} -- 
We first briefly discuss the material setup and, then, devote our main effort to discussing the effect of the $p_{z}$ orbital in these systems. 
There are two ways to overlay thin films on substates, i.e., via van der Waals interactions and via chemical bonding.
The latter one may strongly modify the low-energy electronic structure of the thin films, which can either devastate the topological nature or enhance the nontrivial gap. The second scenario was textbook-likely demonstrated in the bistmuthene/SiC work. 

Here, we propose two silica polymorphs as plausible substrates for III-V honeycomb monolayers, which are widely-used traditional substrates taking advantages of excellent insulation, chemical stability and compatibility with silicon. Specifically, we suggest synthesizing III-V films on $\alpha$-quartz (denoted as $\alpha$-SiO$_2$) and $\beta$-tridymite (denoted as $\beta$-SiO$_2$) silicon dioxides.
Their experimentally determined lattice constant are 4.921~\AA~\cite{d'Amour:a17323} and 5.063~\AA~\cite{doi:10.1080/10408430701718914}, respectively. The equilibrium lattice constants of the three paradigm III-V films studied in this work are 4.805 \AA~(InBi), 4.928 \AA~(TlBi)~\cite{doi:10.1021/nl500206u} and 4.81\AA~(TlSb)~\cite{Li2015}, rendering strain within a range of -0.14\% to 5.37\%, which are fairly reasonable values in experiments. All the six monolayer-substrate systems support room-temperature QSH states with gaps over 200 meV. With $\alpha$-SiO$_2$, the substrate is inert and interacts with III-V monolayers through Van der Waals force, retaining a topological gap of $564$ meV in TlBi. While with $\beta$-SiO$_2$, the $p_{z}$ orbital of the III-V film is bonded to the substrate, rendering a gap up to 907 meV. 

In the geometry with $\alpha$-SiO$_2$ (Fig.~\ref{Fig1}(a)), the III-V film is constructed on the (001)-plane of the substrate with the center of the hexagonal cell above the top silicon atoms. The dangling-bonds of the substrate are saturated by fluorination, in simulation of an experimentally passivated/terminated surface. 
Hence the III-V film does not directly bond with $\alpha$-SiO$_2$ but rather interacts weakly through Van der Waals forces. 
The substrate hardly affects the electronic structure of III-V, leaving the bands around Fermi level similar to the free-standing III-V films (Fig.~\ref{Fig1}(b)). By inspecting orbital characteristics (Fig.~\ref{Fig1}(c, d)), we show that the band degeneracy of $\vert p^+_x\rangle$ and $\vert p^+_y\rangle$ is lifted by the SOC, inducing an indirect gap of 564 meV. The conduction and valance bands show detectable Rashba splitting, primarily induced by the lower $\vert p^-_z\rangle$ state, as will be explained in latter model discussion.   

In the geometry with $\beta$-SiO$_2$ (Fig.~\ref{Fig1}(f)), the III-V film is constructed on the (001)-plane of SiO$_2$ with group-V atoms above the top oxygen atoms. 
In this case, the III-V film is self-assembled and chemically bonds with the substrate. The strong bonding significantly alters the band structure of the film via an orbital filtering effect. 
The $\vert p^-_z\rangle$ orbitals of III-V monolayer are saturated by bonding to the oxygen dangling bonds, and the bands around the Fermi level merely come from $\vert p^+_x\rangle$ and $\vert p^+_y\rangle$ (Fig.~\ref{Fig1}(h, i)). 
They do not display any sizable Rashba splitting. 
As compared to III-V-monolayer/$\alpha$-SiO$_2$, the SOC-induced gap in III-V-monolayer/$\beta$-SiO$_2$ is significantly enhanced to 907 meV in TlBi, larger than in bismuthene/SiC.

We further verify that the nontrivial bulk bands in both systems herald the presence of edge modes inside the energy gap. As attested in Fig.~\ref{Fig1}(e, j), the III-V films combined with the two substrates host topologically nontrivial QSH edge states. The flexibility in overlaying topological monolayers and the relatively easy preparation of a large-scale smooth surface of SiO$_{2}$ reduce the challenge in experimental fabrication and may provide III-V-monolayer/SiO$_{2}$ as a second prototype large-gap QSH system.   

{\it Topological origin and role of $p_{z}$} --
After confirming the topological nature and the enhanced bulk gap, we further address the topological origins of III-V-monlayer/SiO$_{2}$ and compare it to bismuthene/SiC(0001). 
In III-V films, the two different elements enrich low-energy $p$ states, i.e., in addition to the $p_{x}$ and $p_{y}$ orbitals actively participating in bismuthene/SiC, the $p_{z}$ orbital contributes equally in III-V-monolayer/SiO$_{2}$. 
Thus, we use all three $p$ orbitals to construct a tight-binding model using Slater-Koster integrals~\cite{PhysRev.94.1498}. 
For simplicity, we first include up to the nearest neighbor hopping terms.
They are sufficient for correctly interpreting the low energy electronic structure and the associated topology.  Longer-range hopping terms will further improve the quality of the model, but they are not crucial for the following analysis (see the Supplementary Material for more information). 

\begin{figure}[t]
\centering
\includegraphics[width=\linewidth]{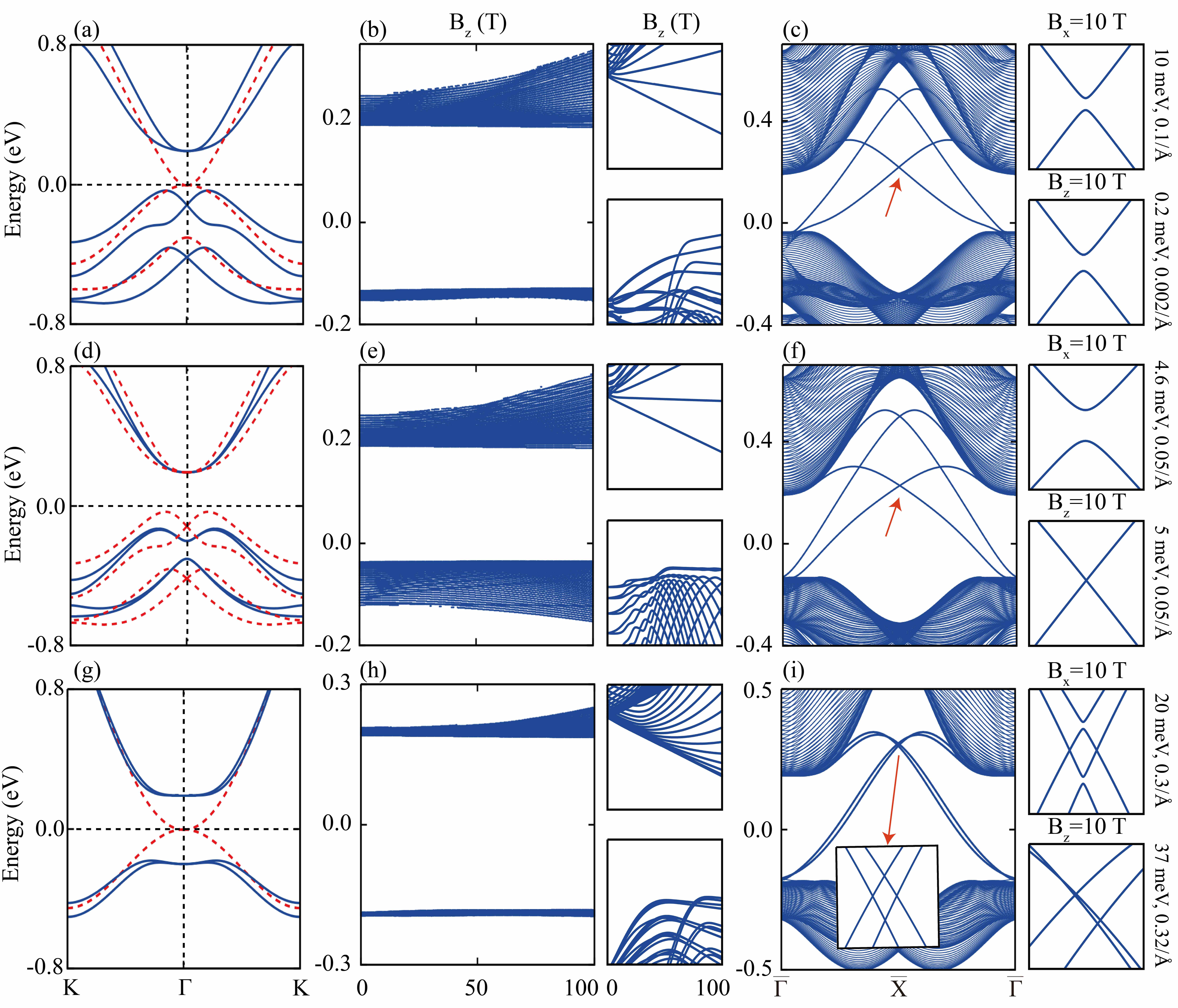}
\caption{(Color online) \textbf{Model Hamiltonian.} (a) Spinless (dashed red line) and spinful (solid blue line) band structures with the basis $p$.  (b) Landau levels as a function of a perpendicular magnetic field, with two zoomed-in plots on the right showing the conduction band bottom and valence band top evolution in an energy-range of $10$ meV. (c) Zigzag edge states calculated from a nanoribbon with bulk bands as (a). (d) Comparison of spinfull band structure (dashed red line) and band structure with vanishing $\langle p_{x/y, \sigma}\vert \bm{L}\cdot \bm{S}\vert p_{z, \sigma}\rangle$ (solid blue line). (e, f) The same as (b, c), but with vanishing $\langle p_{x/y, \sigma}\vert \bm{L}\cdot \bm{S}\vert p_{z, \sigma}\rangle$. (g-i) The same as (a-c), but with the basis $p_{x, y}$. The six zoomed-in plots on the r.h.s correspond to the edge Dirac points denoted by the red arrows in (c, f, i) under magnetic fields $B_x=10$ T ($B_z=10$ T). They are gapped as long as $p_{z}$ orbital is active. The energy and momentum ranges of these subplots are shown on the right of each.}
\label{Fig2}
\end{figure}

Without the SOC, the Hamiltonian is a $6\times6$ matrix spanning on the orbital and site basis $(p_{x}, p_{y},p_{z})\otimes(A, B)$. 
The III-V films are buckled and there exists a small but non-negligible hopping along the perpendicular direction between the A-B sublattices of III-V monolayers on both $\alpha$- and $\beta$-SiO$_{2}$ substrates, which is absent in bismuthene. 
The buckling structure is also consistent with the presence of the $p_{z}$ orbital in the low-energy electronic structure.  
We note that such a $6\times6$ matrix contains redundant information. 
The low-energy electronic structures of the III-V film on SiO$_{2}$ are mainly dominated by three bands, with the other three eigenvalues of the $6\times6$ Hamiltonian lying at higher energy. 
The $p$ orbitals from A and B sites form bonding/anti-bonding states $p^{+}_{\alpha,\sigma}/p^{-}_{\alpha,\sigma}$,
where $\alpha$ denotes the three component of $p$ orbital, and $\sigma$ denotes the spin. 
The local $C_{3v}$ symmetry requires $p^{+}_{x, \sigma}$ ($p^{-}_{x,\sigma}$) and $p^{+}_{y, \sigma}$ ($p^{-}_{y,\sigma}$) to be degenerate at the $\Gamma$ point in the spinless case. 
According to the DFT guidance, we choose a group of parameters setting the $p^{+}_{x, \sigma}$/$p^{+}_{y,\sigma}$ degeneracy at the Fermi level and the $p^{-}_{z}$ band lower in energy. The three low-energy bands are shown in the dashed red line in Fig.~\ref{Fig2}(a). 

Next, we include the atomic SOC $\lambda\boldsymbol{L}\cdot\boldsymbol{S}$ contribution to build up a relativistic analysis. 
The atomic SOC acting on three $p$ orbitals connects $p_{x,\sigma}$ and $p_{y,\sigma}$ by $L_{z}S_{z}$, $p_{x/y,\sigma}$ and $p_{z,-\sigma}$ by $L^{\pm}$. 
The former term is the dominant trigger of topological gap between $p_{x/y,\sigma}$, and the latter one induces the Rashba splitting. 
In bismuthene/SiC(0001), the Rashba spiltting of bands at $K$ is solely due to the presence of the SiC substrate. 

In III-V-monolayer/SiO$_{2}$, the presence of the $p_{z}$ orbital provides a new source of band Rashba splitting. The bands of the $12\times12$ relativistic model are shown in the solid blue line in Fig.~\ref{Fig2}(a), qualitatively resembling the case of III-V-monolayer/$\alpha$-SiO$_2$ (Fig.~\ref{Fig1}(b)). By constructing a nanoribbon, we find extra states connecting the conduction and valence bands inside the bulk gap (Fig.~\ref{Fig2}(c)), evidencing the hallmark of topological QSH edge states. In addition, the bands in Fig.~\ref{Fig2}(a) manifest clear Rashba splitting since the nonequivalent energy levels of the two III- and V-group elements break the inversion symmetry even without explicitly considering the substrate in our model. 

After having convincingly shown that the tight-binding model captures all the essential features, we exploit the model to resolve contributing factors of the above SOC-induced topological gaps in III-V honeycomb lattice. 
We first dissemble the $\sigma \sigma$ and $\sigma \pi$ SOC contributions to examine their effects. 
By retaining term $\langle p_{x, \sigma}\vert \bm{L}\cdot \bm{S} \vert p_{y, \sigma}\rangle$ but switching off $\langle p_{x/y, \sigma}\vert \bm{L}\cdot \bm{S}\vert p_{z, \sigma}\rangle$, we obtain a model conserving $S_{z}$. 
The band structure displays a band gap (the solid blue lines in Fig.~\ref{Fig2}(d)) triggered by the SOC between $p_{x/y}$ as in the case of bismuthene/SiC.
Moreover, the Rashba band splitting is significantly weakened, which reduces the valence band maximum. 
In consequence, the indirect topological gap is increased by an energy scale comparable to the Rashba splitting (comparing the solid blue lines and dashed red lines in Fig.~\ref{Fig2}(d)). 
This indicates that, although it lies completely below the valence bands, the presence of $p_{z}$ orbital in III-V-monolayer/SiO$_{2}$ inevitably reduces the topological gap through the SOC with low-energy $p_{x/y}$ orbitals. 

The negative effect on gap size made by $p_z$ can apparently be eliminated if the orbit is bonded. 
Removing $h_{\alpha z}^{ij}$ from the Hamiltonian we reduce the model to $4\times 4$ and $8\times 8$ matrices in basis of $(p_{x}, p_{y})\otimes(A, B)$ for spinless and spinful cases respectively. 
The band structures and corresponding edge states with the same parameters as in Fig.~\ref{Fig2}(a, c) are shown in Fig.~\ref{Fig2}(g, i). Directly corresponding to III-V-monolayer/$\beta$-SiO$_2$, the SOC-induced topological gap is substantially magnified in the absence of $p_z$. 
This may accomplish another improvement over bismuthene/SiC. 
In both III-V-monolayer/$\beta$-SiO$_{2}$ and bismuthene/SiC, the low-energy physics is govern by $p_{x/y}$ orbitals.
They both show sizable band splitting, derived from either the inequivalent sublattices or the substrates. 
However, the band splitting is stronger at $K$ and weaker at $\Gamma$ in both systems. 
As a result, the gap size of III-V-monolayer/$\beta$-SiO$_{2}$ is hardly affected by the Rashba splitting. 
In this respect, for QSH states promoted by $p_{x/y}$ orbitals on honeycomb lattices, the largest topological gap can be potentially realized by the mechanism of band-inversion at $\Gamma$. This also explains why TlBi/$\beta$-SiO$_2$ harbors an even larger QSH gap than bismuthene/SiC does, when Tl has a lower atomic number than Bi.

\begin{figure}[htbp]
\centering
\includegraphics[width=\linewidth]{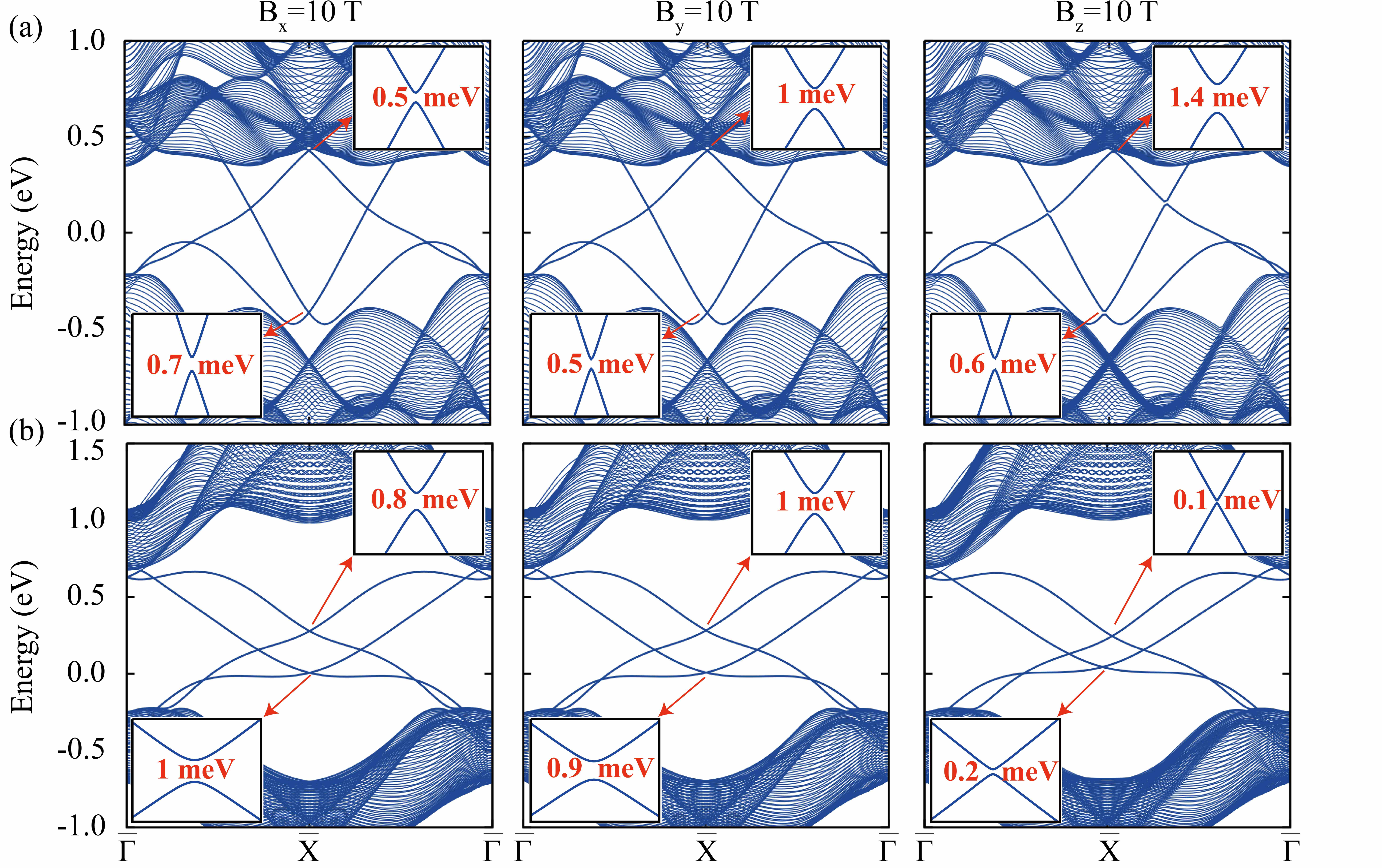}
\caption{(Color online) \textbf{Magnetic responses of III-V-monolayer/SiO$_2$.} (a) TlBi/$\alpha$-SiO$_2$. (b) TlBi/$\beta$-SiO$_2$. The insets are zoomed-in plots in an energy-range of $5$ meV and a $k$-range of $0.05$/\AA.}
\label{Fig3}
\end{figure}

{\it Magnetic response} -- 
After understanding the positive role of $p_{z}$ on the topological gap size, next we examine its role on the magnetic response. 
In a band-inversion type QSH system, the robustness of the nontrivial topology is controlled by the gap size at $\Gamma$ instead of the indirect gap~\cite{PhysRevLett.114.096802, PhysRevB.90.115305}. 
Thus, despite the small global gap, the quantized conductance inherent from the QSH states in a InAs/GaSb quantum-well persists to a large magnetic fields in experiment, indicating that the breaking of time reversal symmetry does not lead to an immediate trivialization of the system. 
In III-V-monolayer/SiO$_{2}$, the inverted band gap at $\Gamma$ is much more pronounced, which is expected to promise a significant stabilization of the QSH states against external magnetic field. 
Furthermore, as long as $S_{z}$ is conserved, the topological edge states remain robust under a perpendicular magnetic field. 
However, due to the presence of either the sublattice and/or the substates, the $p_{z}$ orbital will mix with $p_{x/y}$ orbitals through SOC,  breaking $S_{z}$ conservation. 
Consequently, the topological edge states may become no longer stable.    

To test the robustness of QSH states in III-V-monolayer/SiO$_{2}$ and further elaborate the role of $p_{z}$, we study the magnetic response under both a longitudinal and transversal field.  
 We simulate the developments of the Landau Levels (LLs) under a perpendicular magnetic field $B_z$ by introducing a Peierls phase in the model Hamiltonian of nanoribbon. As shown in Fig.~\ref{Fig2} (b, e, h), with the increase of $B_z$, the bulk LLs of the conduction/valence bands hardly change, and they do not cross under any experimentally accessible strength of the magnetic field. 
Thus, owing to the large topological gap the bulk band topology is much more robust than n HgTe and in InAs/GaSb. 
One cannot destroy the QSH states by only breaking the time reversal symmetry in III-V-monolayer/SiO$_{2}$. 

However, the edge states can be gapped in presence of the $p_{z}$ orbital and the breaking of inversion symmetry.
Note that, under a perpendicular magnetic field, $B_z$ mixing opposite helicities. Edge states are not severely destroyed and they may persist nearly gapless, as exemplified by the inversion symmetric HgTe quantum wells~\cite{PhysRevB.82.155310, PhysRevB.86.075118, PhysRevB.98.161407, Rothe_2010, PhysRevLett.109.216602, PhysRevB.85.125401, PhysRevB.86.075418, PhysRevB.91.235433, PhysRevB.94.035146, PhysRevLett.104.166803, TKACHOV2012900, PhysRevB.90.115305}.
We find that, for III-V films with broken inversion symmetry, the gaps of the topological edge states are always negligibly small although the Rashba band splitting is fairly large. 
The six subplots on the r.h.s. of Fig.~\ref{Fig2} are zoomed-in plots of edge Dirac points denoted by the red arrows in Fig.~\ref{Fig2}(c, f, i) under $B_{x}$ and $B_{z}$, correspondingly. 
With all three $p$ orbitals as shown in Fig.~\ref{Fig2}({c}), the edge states are gapped under both in-plane and out-of-plane magnetic fields, but only in the order of meV. While, with vanishing SOC coupling between $p_{x/y}$ and $p_{z}$ (Fig.~\ref{Fig2}(f)) or with only $p_{x/y}$ orbitals (Fig.~\ref{Fig2}(i)), the edge Dirac points in this inversion asymmetric environment remain gapless under $B_z$.
Going from the simplified model to the full DFT band structure, after taking into account all sublattice asymmetry and the substrate effect, we find the gap on the edge states induced by the external magnetic field and the breaking of inversion symmetry remains tiny in both systems.
The largest gap resolved in our calculations is only $1.4$ meV under $B_{z} = 10~T$, as shown in Fig.~\ref{Fig3}.  
Bonding of $p_{z}$ with $\beta$-SiO$_{2}$ further eliminates part of Rashba contributions, leading to an even smaller gap between the edges states under $B_{z}$. 
Thus, a conductance measurement of III-V-monolayer/SiO$_{2}$ may observe a more robust quantized plateau (in large-range of experimentally accessible magnetic fields $B_{z}$) than in HgTe, which is highly promising for device application. 

{\it Conclusion.} -- We offer a general band-inversion model Hamiltonian of planar/buckled hexagonal sheets constituted by $p$-orbitals. Base on the model, an emerging QSH state from gapping the degenerate bonding/antibonding states of $p_{x, y}$ is explained. The off-diagonal SOC sector, hybridizing opposite spins and sublattice degrees of freedom, is of crucial importance. It, on one hand, reduces the topological gap sizes via Rashba splitting, on the other hand, it opens gaps on edge states in perpendicular magnetic fields. The two effects manifest themselves in realistic systems and offer us a tuning factor to engineer promising, application-relevant systems. 

We, accordingly, offer two strategies to boost topological gaps by manipulating $p_z$ orbitals.
One is to engineer the topological mechanism from Kane-Mele type to band-inversion type, such that the Rashba band splitting is significantly reduced at $\Gamma$ as exemplified in $\alpha$-quartz SiO$_2$.
The second strategy takes a further step to directly bond $p_z$ of III-V monolayer to $\beta$-tridymite SiO$_2$. This eliminates the above mentioned off-diagonal couplings and promoting the topological gap up to $907$ meV in TlBi, taking the full advantage of the $p_{x/y}$ intrinsic SOC.  
For the same reason, nearly gapless edge states in III-V-monolayer/$\beta$-SiO$_2$  are observed under out-of-plane magnetic fields in the ballistic limit. 

The successful implementation of the band-inversion type topological mechanism in two-dimensional III-V monolayers, and the proposed tight-binding models are amenable to all $p$ hexagonal monolayers, providing a platform for possible future in-depth theoretical and experimental studies. The large gap QSH edge states in the two III-V-monolayer/SiO$_2$ systems are viable for room-temperature fabrication and manipulations. We expect our work to elevate QSH from a low-temperature quantum phenomenon to an applicable effect in modern electronic devices.

{\it Acknowledgements.} -- 
This work was supported by the National Key R$\&$D Program of China (2017YFE0131300), the Strategic Priority Research Program of Chinese Academy of Sciences (Grant No. XDA18010000), Shanghai Technology Innovation Action Plan 2020-Integrated Circuit Technology Support Program (Project No. 20DZ1100605), and Chinese-German Mobility program: M-0006.
W. H and R. C. are also partly supported by the W\"urzburg-Dresden Cluster of Excellence on Complexity and Topology in Quantum Matter-ct.qmat. 
Calculations were carried out at the HPC Platform of ShanghaiTech Laboratory for Topological Physics and School of Physical Science and Technology.



\bibliographystyle{apsrev4-1}
\bibliography{ref}  

\end{document}